# High-reflectivity, high-*Q* micromechanical membranes via guided resonances for enhanced optomechanical coupling

Catvu H. Bui,[1,*] Jiangjun Zheng,[1,*] S. W. Hoch,[2] Lennon Y. T. Lee,[1] J. G. E. Harris,[2,3] and Chee Wei Wong[1]

[1] Optical Nanostructures Laboratory, Columbia University, New York, NY 10027, USA

[2] Department of Physics, Yale University, New Haven, CT 06520, USA

[3] Department of Applied Physics, Yale University, New Haven, CT 06520, USA

Using Fano-type guided resonances (GRs) in photonic crystal (PhC) slab structures, we numerically and experimentally demonstrate optical reflectivity enhancement of high-*Q* $SiN_x$ membrane-type resonators used in membrane-in-the-middle optomechanical (OM) systems. Normal-incidence transmission and mechanical ringdown measurements of 50-nm-thick PhC membranes demonstrate GRs near 1064 nm, leading to a ~ 4× increase in reflectivity while preserving high mechanical *Q* factors of up to ~ $5 \times 10^6$. The results would allow improvement of membrane-in-the-middle OM systems by virtue of increased OM coupling, presenting a path towards ground state cooling of such a membrane and observations of related quantum effects.

[*] Equal contribution.
Electronic addresses: chb2127@columbia.edu and jz2356@columbia.edu

Among current experimental setups in cavity optomechanics [1,2], the membrane-in-the-middle configuration permits a quadratic dependence of cavity resonances $\omega_{cav}(x)$ on membrane displacement $x$, allowing the possibility of quantum non-demolition readout of the membrane's phonon number [3-6]. The curvature of the detuning $\omega_{cav}(x)$ determines the strength of the OM coupling relevant for this type of readout. When the membrane is placed at a node or antinode of the intracavity mode, this curvature depends on the reflectivity $|r_m|^2$ of the membrane, which must approach ~ 0.99 for quantum effects to be detectable [3]. Although OM effects such as strong laser cooling have been demonstrated in this platform [3], the 50-nm-thick $SiN_x$ membrane's low $|r_m|^2$ of ~ 0.13 represents an important technical challenge. The difficulty comes from the fact that much of this platform's performance is afforded by the remarkably high mechanical $Q$ factor [3,7,8], low motional mass and low optical absorption of the $SiN_x$ membrane, which would be compromised by conventional reflectivity-enhancing solutions such as multilayered Bragg mirrors or metal coatings. So far, high-$Q$ and low-mass thin mirrors developed for optomechanics [9-12] have not achieved comparable motional mass and $Q$ factors realized in the 50-nm-thick $SiN_x$ micromechanical membranes described in Ref. [3,7,8]. There are also other approaches that seek to increase the curvature of the detuning $\omega_{cav}(x)$ via avoided crossings between intracavity modes, hence circumventing the need for highly reflective resonators [5,6]. Progress towards reaching the quantum regime is, however, still ongoing.

Motivated by this challenge, we demonstrate Fano-type GRs in PhC slab structures as an alternative means to optimize OM $SiN_x$ membranes for enhanced reflectivity. The phenomenon of GRs, where the interference coupling between discrete, leaky crystal modes and the continuum of free-space radiation modes gives rise to asymmetrical Fano-like line shapes on the reflection and transmission spectra [13-16], has been widely studied due to its potentials for compact sensing and communication device applications [17-23]. Here we demonstrate its application in cavity optomechanics by patterning small-area hole arrays into high-$Q$ 50-nm-thick $SiN_x$ membranes as a reflectivity-enhancing modification. Ab-initio simulations and transmission measurements verify GRs and transmission dips near 1064 nm, which is the laser wavelength used in Ref. [3] to excite the Fabry-Perot cavity in the membrane-in-the-middle setup. More importantly, the membranes' high $Q$ factors are confirmed (by mechanical ringdown measurements) to be preserved in the presence of the PhC structures, thus ensuring an improved OM performance by realizing in the same device a good reflector and a high $Q$, low mass, ultrathin resonator.

We first design our PhC structures numerically. A square lattice of $C_{4v}$ point group is used for the PhC slab design with a slab thickness $t \sim 50$ nm and a refractive index $n_{SiN} \sim 2.15$. To structurally tune normal-incidence transmission spectral features of the PhC slab, we make use of the dependence of GRs' locations on the normalized thickness $t/a$, where $a$ is the PhC's lattice constant [21]. Transmission spectra for different values of $t/a$ [Fig. 1(a)] are computed using 3D finite-difference time-domain (FDTD) method (RSoft FullWAVE), similar to that in Ref. [13]. As shown in Fig. 1(a-bottom), a PhC slab having $a = 967$ nm, $t = 50$ nm, and a hole diameter $d = 290$ nm, corresponding to $t/a = 0.052$ and $d/a = 0.3$, is found to produce a transmission dip at $\lambda \sim 1064$ nm. Band structure of this PhC slab [Fig. 1(b)] is also computed using plane wave expansion method (Rsoft BandSOLVE) and shows, at the Γ point, a doubly degenerate mode at normalized frequency $0.896(2\pi c/a)$ [mode $A$ in Fig. 1(b)]. This leaky mode manifests itself on the transmission spectrum in Fig. 1(a-bottom) as the dip at 1064 nm. Singly degenerate modes, such as mode $B$ in Fig. 1(b), are unable to couple to normal-incidence illumination due to symmetry mismatch [13,14,24] and, as we will see later, only show up on the transmission spectrum under non-normal illumination.

We fabricate our samples according to structural parameters described above, starting with commercial low-stress $SiN_x$ membranes (Norcada Inc.) of nominal dimensions of 1 mm × 1 mm × 50 nm, suspended from 200-μm-thick silicon frames [Fig. 2(b)]. For compatibility with spin-coating and vacuum processes, the frame is temporarily attached to a silicon substrate pre-etched with small pressure-regulating venting trenches [Fig. 2(a-I)]. Electron-beam lithography, with PMMA as resist, is used to define the hole array patterns [Fig. 2(a-II)]. After development in MIBK:IPA solution, the pattern is transferred to the $SiN_x$ layer using an optimized $CF_4/O_2$ RIE etch [Fig. 2(a-III)]. Finally, the PMMA is removed with acetone [Fig. 2(a-IV)]. SEM images [Fig. 2(c)] demonstrate the fabrication quality. The hole array consists of 250 × 250 holes at the center of the membrane, covering the $1/e^2$ diameter of the Nd:YAG laser's $TEM_{00}$ cavity mode (90 μm) [3] while minimizing the PhC area to only ~ 6% of the membrane area to preserve its mechanical properties.

To verify mechanical $Q$ factors of the membranes in the presence of PhC structures, mechanical ringdown measurements are performed on the patterned membranes at $10^{-6}$ Pa and room temperature using a method similar to that of Ref. [7]. Measured resonant frequencies and associated $Q$ factors for several vibrational modes are plotted in Fig. 3(a). A typical ringdown is shown in Fig. 3(b). These results show $Q$ factors to remain at high values of up to ~ $5 \times 10^6$, similar to those reported for unpatterned membranes [3,7,8], thus establishing that introducing the hole array indeed does not appreciably degrade the membrane's high $Q$ factors.

Broadband normal-incidence transmission measurement is performed on the fabricated PhC membrane [Fig. 4(a)]. The experimental setup is shown in the inset of Fig. 4(a). The transmission measurement is repeated for different polarizer angles $\theta$ ranging from $0°$ to $90°$ [Fig. 4(b)]. In all cases, we observe strong transmission dips (ranging from 43% to 59%, with associated linewidths ranging from 3.1 nm to 6.5 nm) at ~ 1035 nm, indicating enhanced reflectivity over the membrane's spectral background. Similar to previous reports, the spectra are found to be relatively polarization-insensitive [19,20,22]. The slight wavelength deviation from FDTD simulation could be explained by fabrication-related disorder. We also note that the suppression of transmission could be made broadband [20], thus lessening the requirement of having a GR at exactly 1064 nm. Transmission spectra are also measured at different angles of incidence with fixed polarization [Fig. 4(c)]. When the angle of incidence is varied from $0°$ to ~ $5°$, the dip at ~ 1035 nm become less pronounced and another dip appears at ~ 1070 nm. This represents the coupling of non-normal components of illumination to a singly degenerate mode, specifically to mode $B$ in the band structure as shown earlier in Fig. 1(b) [13,14,24]. It should be noted that the performance observed here is most likely limited by fabrication-related disorder. Since various types of structural disorder affect the GRs' spectra differently [18,20,26], we instead quantify losses from disorder by using a theoretical model for a single GR, first described in Ref. [13] and modified in Ref. [17] to account for losses. We use the term $\tau/\tau_{\text{loss}}$ in such model to quantify losses, where $1/\tau$ is the decay rate of the GR and $1/\tau_{\text{loss}}$ is the extra decay term associated with all losses caused by fabrication-related disorder. We estimate $\tau/\tau_{\text{loss}}$ to be ~ 1 for the spectral performance observed here. For quantum effects to be detectable, membrane reflectivity needs to approach ~ 0.99, which is estimated to correspond to $\tau/\tau_{\text{loss}}$ approaching ~

0.01. While a challenging, as fabrication becomes optimized, this requirement should be achievable.

In the context of optomechanics, these results translate to a membrane of enhanced reflectivity of up to 57 % at the target wavelength, assuming a negligible absorption of the ultrathin SiN$_x$ membrane [3,5]. The background of the spectrum in Fig. 4(a) corresponds to the reflectivity of an unpatterned SiN$_x$ membrane [13] and is ~ 15 %, similar to reported values [3,6]. This work thus represents a ~ 4× increase in reflectivity. To quantify the improvement on OM performance, we calculate and plot in Fig. 4(d) the expected cavity detuning for a cavity containing such a membrane as a function of membrane position $x$, given by [4]

$$\omega_{\text{cav}}(x) = c\phi_r / L + (c/L)\cos^{-1}\left[|r_m|\cos^{-1}(4\pi x/\lambda)\right] \quad (1)$$

where the wavelength $\lambda = 1064$ nm and cavity length $L = 6.313$ cm with a free spectral range FSR of 2.374 GHz. $\phi_r$ is the complex phase of $r_m$. As mentioned, OM coupling strength increases with the curvature of $\omega_{\text{cav}}(x)$. For a patterned membrane of $|r_m|^2 = 57\%$, the detuning curve achieves $\omega''_{\text{cav}}(x)/2\pi = 108$ kHz nm$^{-2}$ at its extrema, a marked improvement over $\omega''_{\text{cav}}(x)/2\pi = 42$ kHz nm$^{-2}$ for an unpatterned membrane of $|r_m|^2 = 15\%$. Such enhanced cavity detuning [Fig. 4(d)] also shows an improvement in the traditional linear coupling regime when the membrane is not placed at a node or antinode of the intracavity mode.

In this work, we demonstrate enhanced reflectivity and preserved high mechanical $Q$ factors, typically in excess of $1 \times 10^6$ with the ($Q_{i,j} \times \nu_{i,j}$) product reaching $3.3 \times 10^{12}$ Hz, of SiN$_x$ membranes. The enhanced reflectivity of the low mass, low loss membranes should allow significantly improved OM coupling via increased radiation pressure per photon, presenting a path for membrane-in-the-middle type OM systems to achieve ground state cooling and observations of single phonon quantum jumps.

The authors acknowledge A. Jayich, J. Gao, and H. Hosseinkhannazer for helpful discussions, and funding support from the DARPA ORCHID program through a grant from AFOSR (J. Z., S. W. H., J. G. E. H., and C. W. W.), NSF 0855455 (S. W. H., J. G. E. H.), DARPA Young Faculty Award (S. W. H., J. G. E. H.), and AFOSR FA9550-90-1-0484 (S. W. H., J. G. E. H.). The Columbia cleanroom facilities are supported by NSF Award CHE-0641523 and by NYSTAR.

**Figure Captions**

FIG. 1. (Color online) (a) Normal-incidence transmission spectra of PhC slabs ($n_{SiN}$ = 2.15 and $d/a$ = 0.3) with $t/a$ = 0.3, 0.1, and 0.052 (top to bottom, respectively). Fixing $t$ = 50 nm, panel (a-bottom) corresponds to a slab having $a$ = 967 nm, $d$ = 290 nm, and producing a dip (red dotted rectangular) at ~ 1064 nm. (b) Corresponding partial band structure of panel (a-bottom), along with electric field intensity distributions of leaky modes at $\Gamma$ point: doubly degenerate mode $A$ at 0.896 ($2\pi c/a$) and singly degenerate mode $B$ at 0.886 ($2\pi c/a$). Mode A appears in panel (a-bottom) as the transmission dip.

FIG. 2. (Color online) (a) Fabrication process: (a-I) Membrane frame attached to substrate; (a-II) PMMA spin-coating and electron beam lithography; (a-III) Pattern transfer using RIE etch; and (a-IV) PMMA and substrate removal. (b) Commercial 1 mm × 1 mm × 50 nm $SiN_x$ membrane suspended from a silicon frame. (c) SEM images showing part of the fabricated hole array at the center of the membrane. Inset shows close-up view of air holes.

FIG. 3. (Color online) (a) Measured $Q$ factors and resonant frequencies at $10^{-6}$ Pa and room temperature of various mechanical modes of the patterned membrane. The membrane is driven at its mechanical resonances by a piezoelectric actuator which is then rapidly switched off. The modes are identified and indexed as $v_{i,j}$ where $i$ and $j$ are positive integers relating the eigenfrequencies as $v_{i,j} = v_{1,1}\sqrt{(i^2+j^2)/2}$, similar to a stressed square membrane. The specific modes are also confirmed by sweeping the laser beam across the membrane's surface and noting the expected locations of vibrational nodes. The modes are carefully driven to avoid the nonlinear Duffing oscillator regime [25]. (b) A typical mechanical ringdown of the PhC

membrane, shown here for mode (3,2) at $v_{3,2}$ = 498 kHz with ~ 3.2 s 1/$e$ lifetime. Solid line is the exponential fit.

FIG. 4 (Color online) (a) Measured normal-incidence transmission spectrum of the fabricated PhC membrane, shown along with FDTD simulation. Inset shows setup: Light from a broadband source (BL) coupled to a multimode fiber is collected by an objective lens (MO) and illuminates the membrane at the PhC site. Transmitted light is collected by a second object lens (MO) and coupled into a single mode fiber connected to an optical spectrum analyzer (OSA). For polarization dependence measurements, a linear polarizer (PL) is placed after the second objective lens with adjustable polarizer angle $\theta$. Incident angle of illumination could be varied by rotating the sample in the direction shown (red arrow). All measurements are normalized to the same measurements with no sample in place. (b) Measured transmission spectra at varied polarizer angles ($\theta$ ~ 0° to 90°) (c) Measured transmission spectra at varied illumination's incidence angles (0° to ~ 5°, in arbitrary steps). Note the coupling to the singly degenerate mode as incidence angle moves away from surface-normal direction. (d) Computed OM detuning curves of the membrane-in-the-middle platform, at normal and enhanced membrane reflectivity. $\omega_{cav}(x)$ is given in units of $2\pi \times$FSR where FSR = 2.374 GHz.

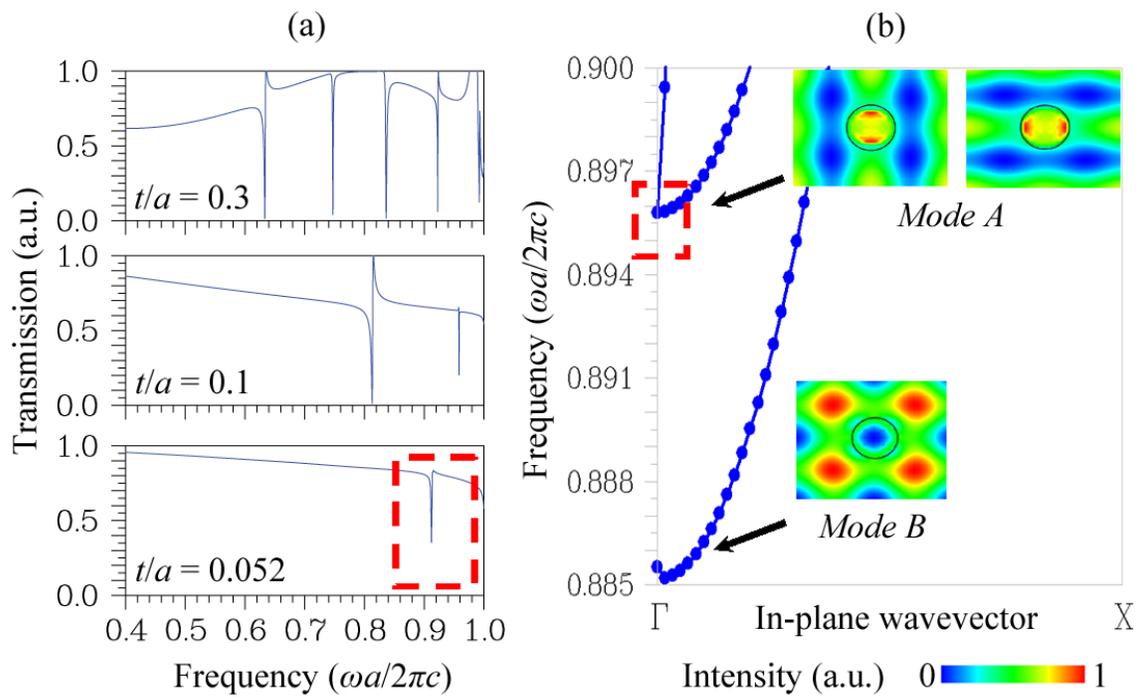

FIG. 1

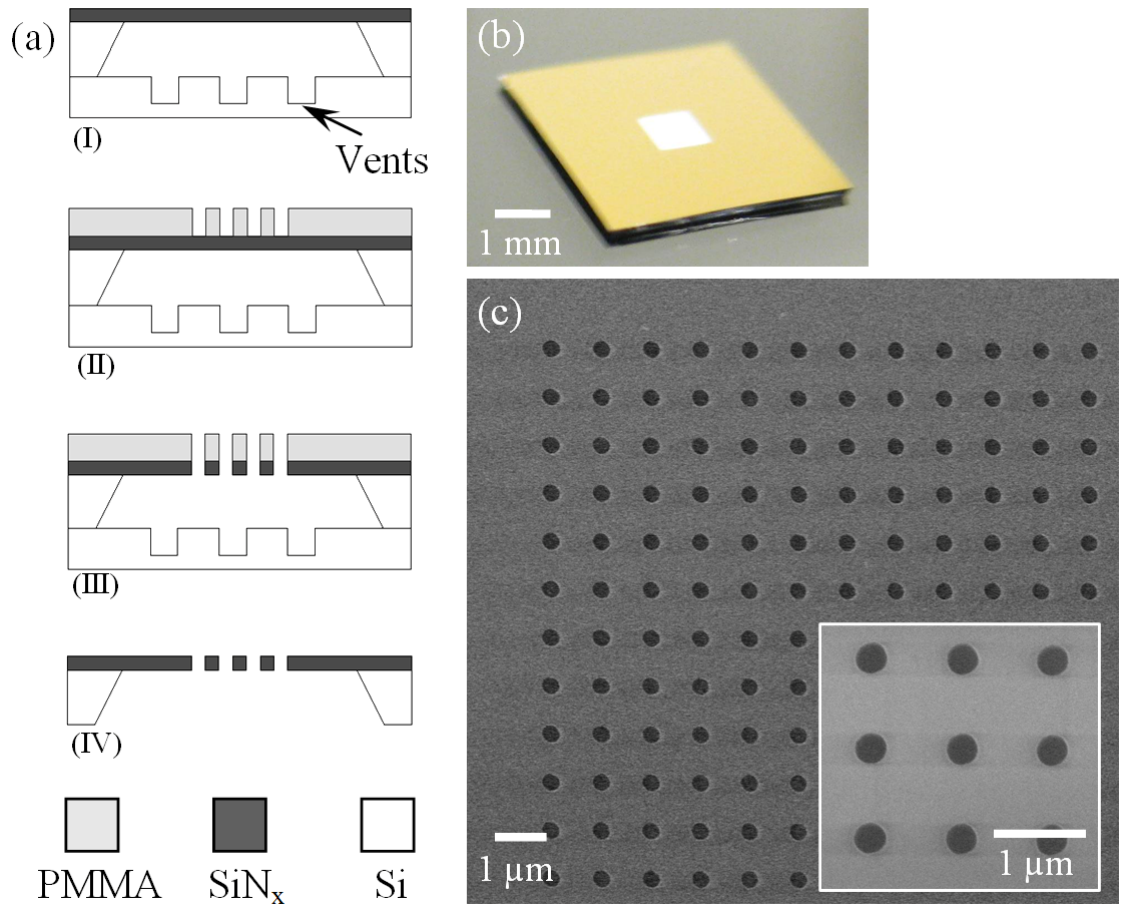

FIG. 2

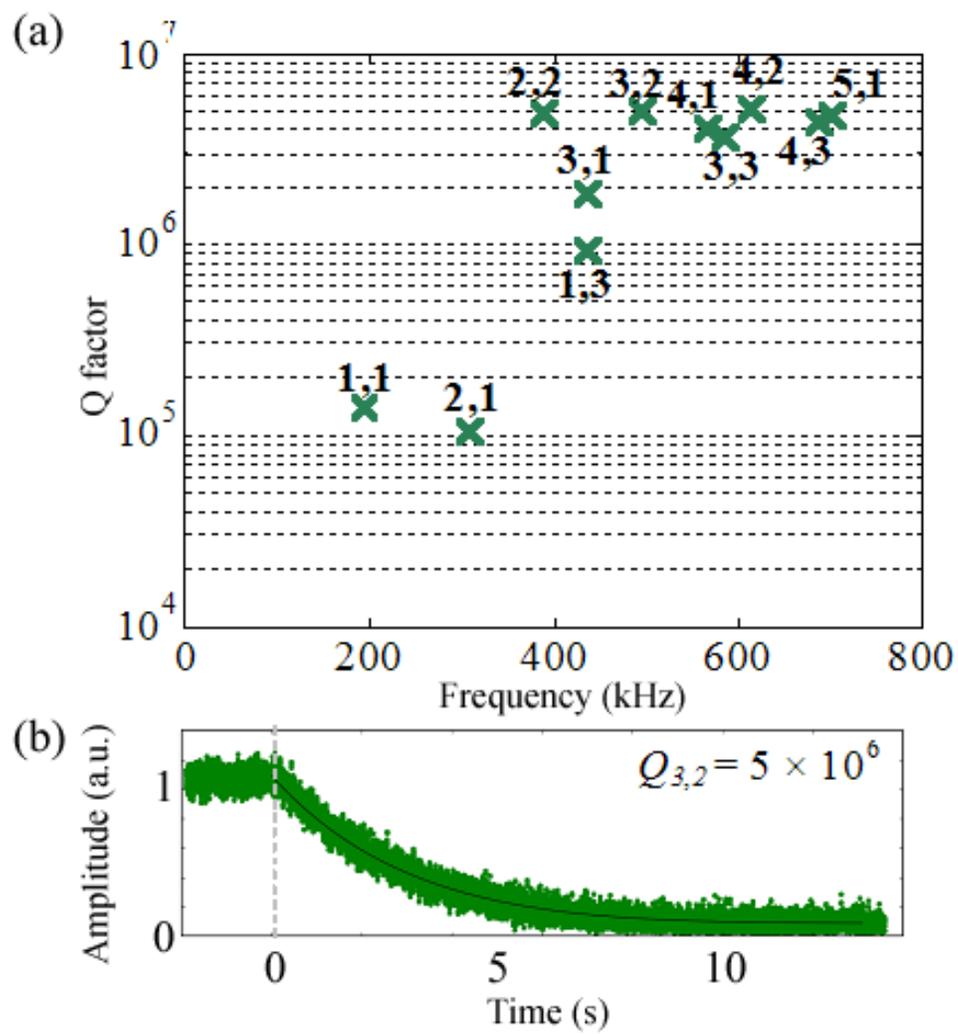

FIG. 3

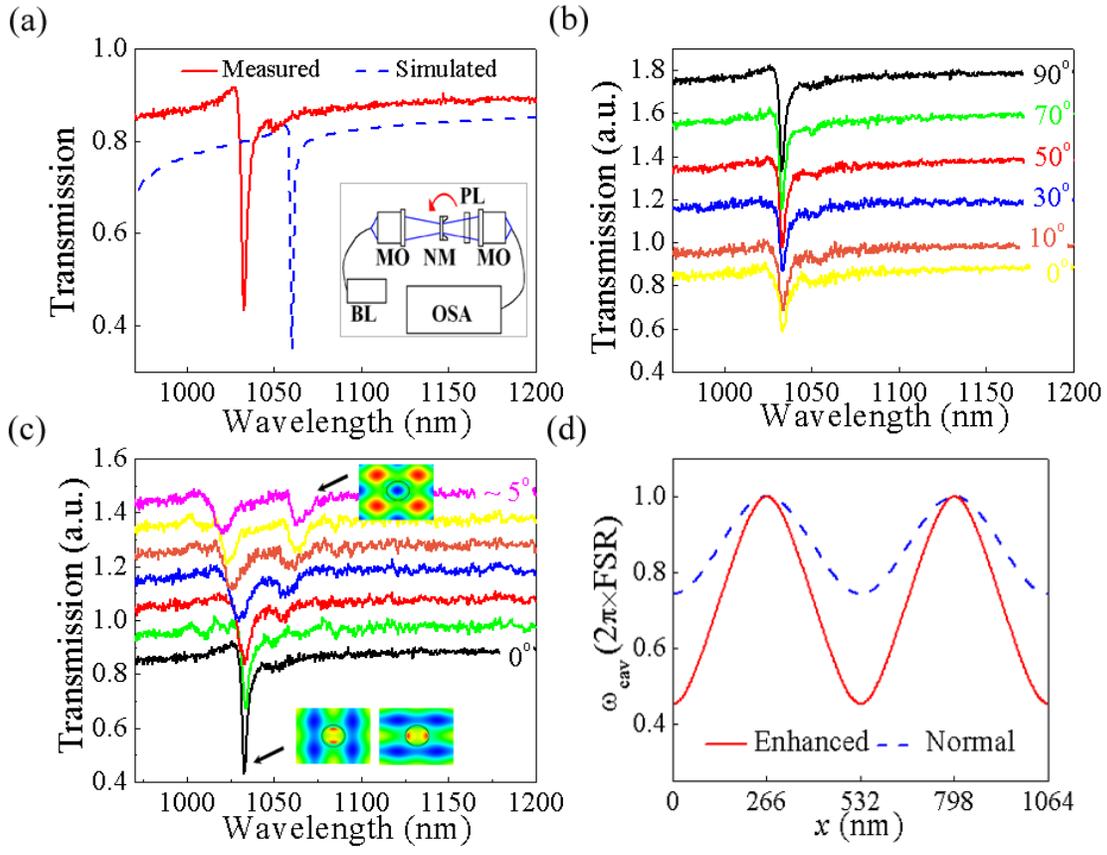

FIG. 4